\documentclass[twocolumn,pre,showpacs]{revtex4}
\usepackage{graphicx,color}
\usepackage{dcolumn}
\usepackage{amsmath,amssymb}
\def\TcScaling{0.907}
\def\L0Scaling{6.5}
\def\TcExtrap{0.905}
\def\figurescale{0.84}
 \def\ZUichi{
 \begin{figure}[t]
  \begin{center}
   \includegraphics[width=\figurescale\linewidth]{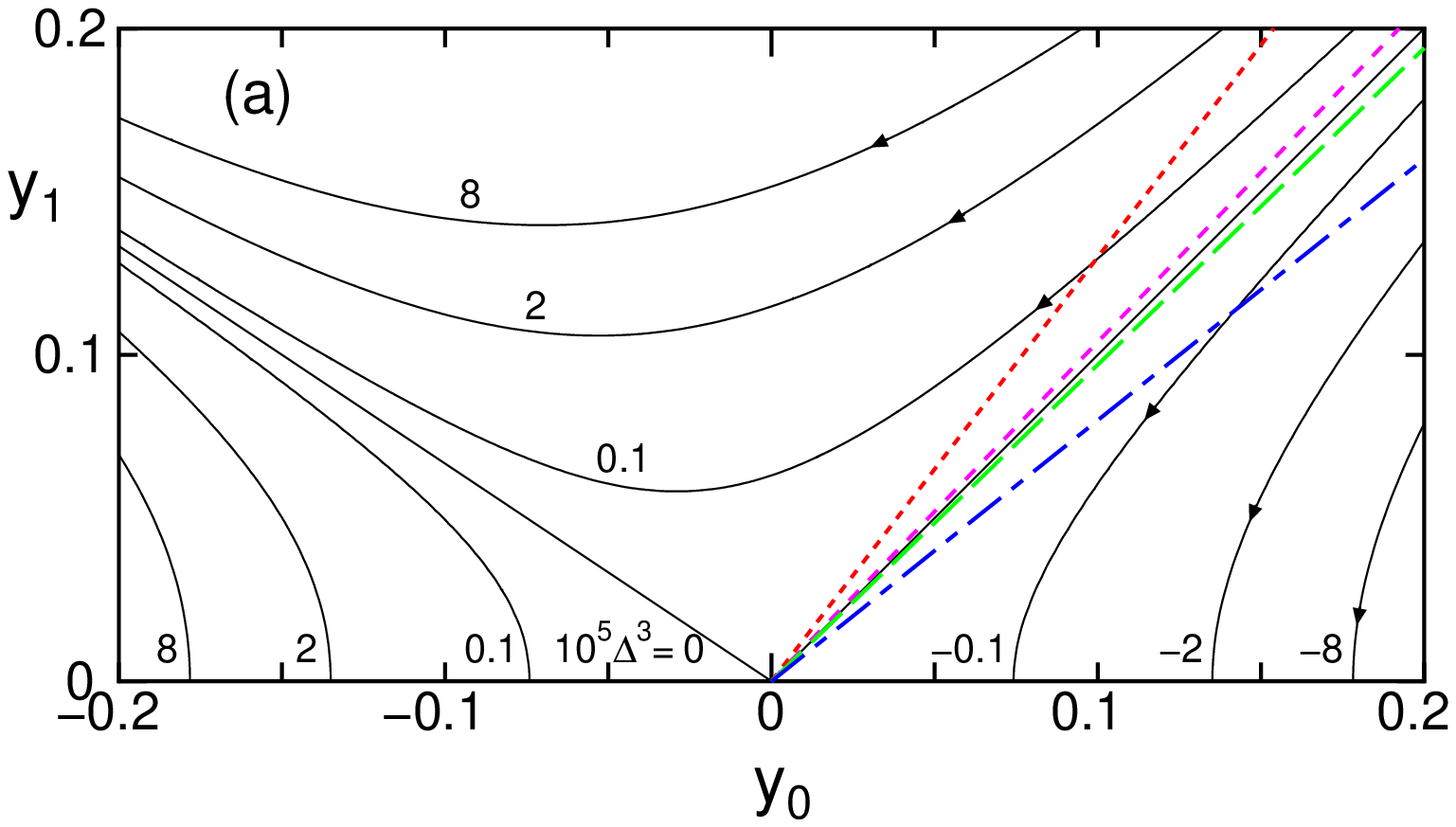}
   \includegraphics[width=\figurescale\linewidth]{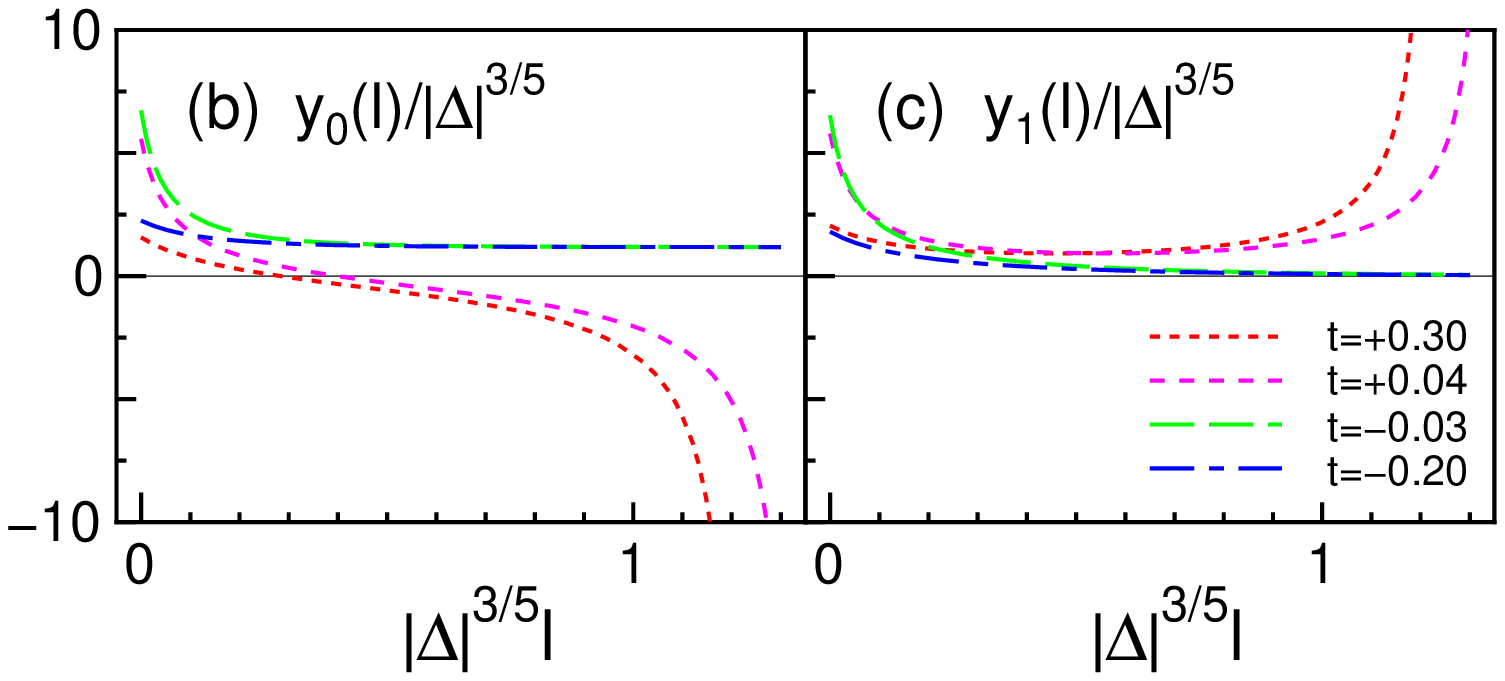}
  \end{center}
  \caption{
  (Color online)
  (a): 
  The contour plot of $\Delta^3$ indicating the RG flow.
  Numerical values near the contour lines in black denote
  $10^5\times\Delta^3$.
  The straight lines in color show $y_1(0)=(1+t)y_0(0)$ for some values
  of $t$:
  0.30 (red dotted line),
  0.04 (pink short-dashed line),
  $-0.03$ (green dashed line),
  and
  $-0.20$ (blue dot-dashed line). 
  (b) and (c):
  The scaled couplings $y_{0,1}/|\Delta|^{3/5}$ are given as functions
  of the scaled variable $|\Delta|^{3/5}l$.
  }
  \label{contour-y0-y1}
 \end{figure}
 }
 \def\ZUni{
 \begin{figure}[t]
  \begin{center}
   \includegraphics[width=\figurescale\linewidth]{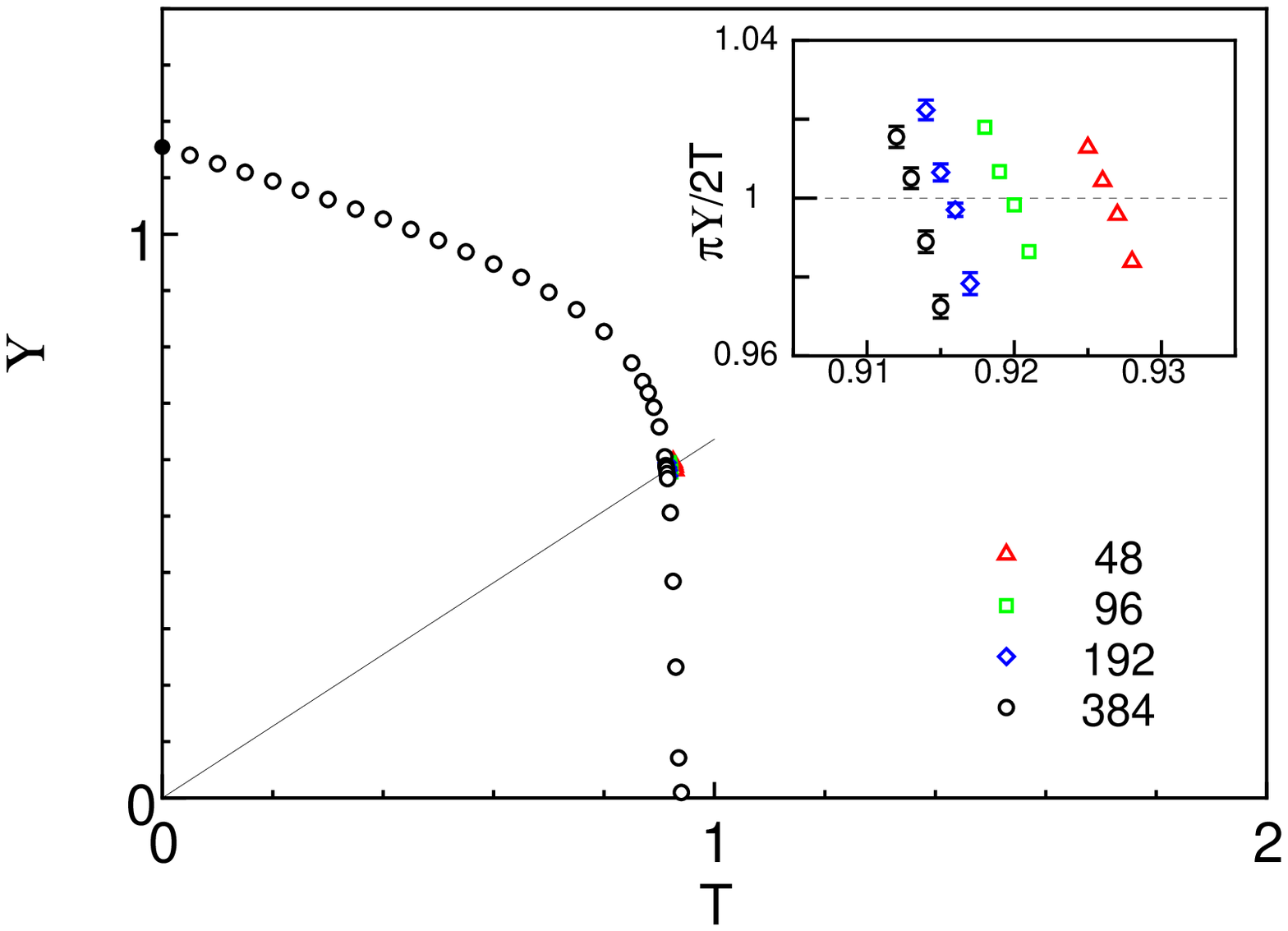}
  \end{center}
  \caption{
  (Color online)
  Temperature dependence of the helicity modulus.
  The solid line exhibits $2T/\pi$, and the filled circle is
  $\Upsilon(0)=\zeta$.
  The inset gives the magnified view near the transition temperature,
  where ${\pi\Upsilon}/{2T}$ is plotted against $T$.
  }
  \label{HelicityModulus}
 \end{figure}
 }
 \def\ZUsan{
 \begin{figure}[t]
  \begin{center}
   \includegraphics[width=\figurescale\linewidth]{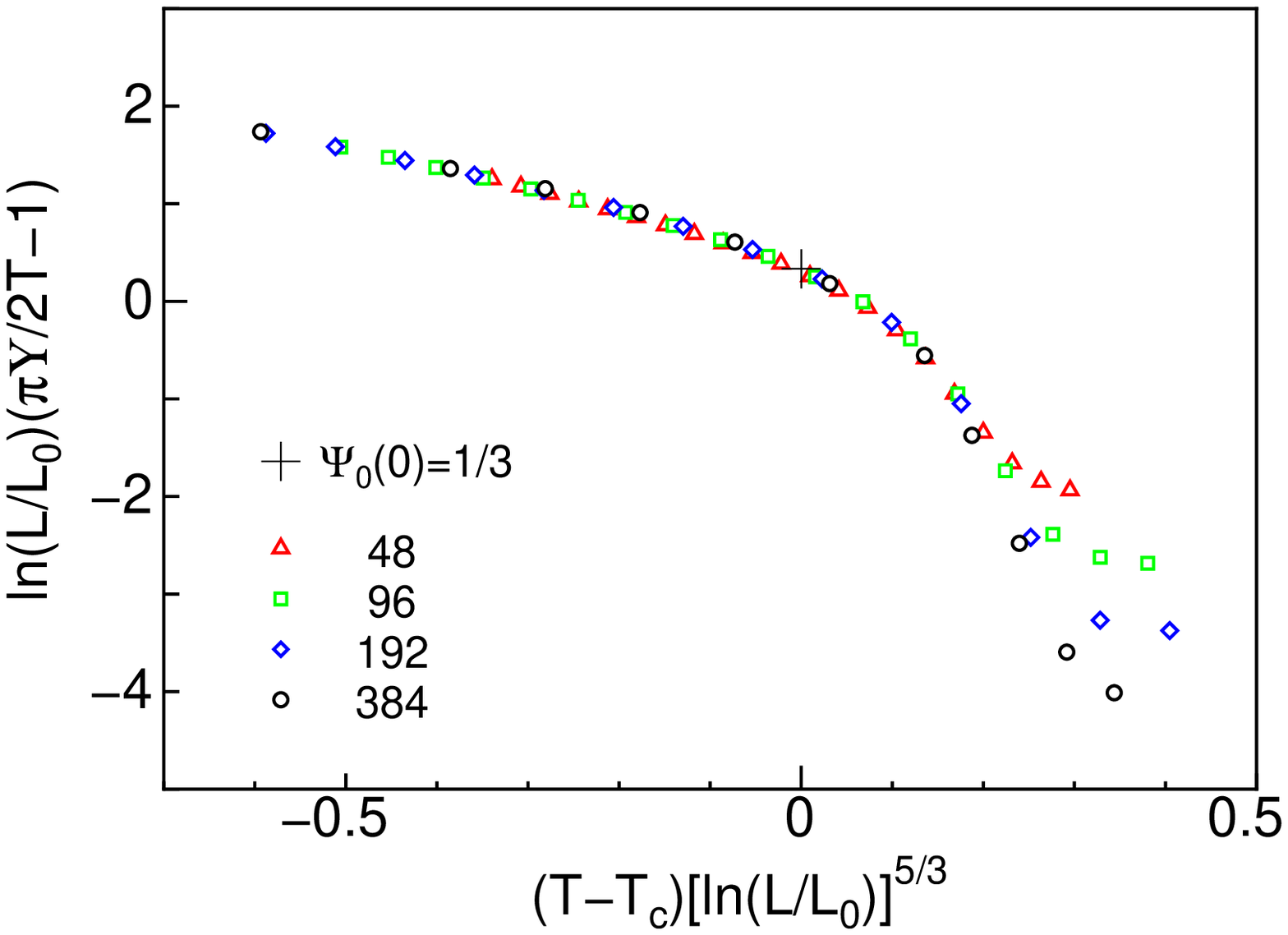}
  \end{center}
  \caption{
  (Color online)
  The FSS plot of the helicity modulus.
  $T_{\rm c}=\TcScaling$ and $L_0=\L0Scaling$ are employed
  according to the best by-eye scaling.
  The cross $+$ shows the theoretical prediction on the scaling function
  $\Psi_0(0)=1/3$ (see the text).
  }
  \label{scaling}
 \end{figure}
 }

\begin{document}
 \title{
 Finite-size-scaling ansatz for the helicity modulus of the triangular-lattice\\
 three-spin interaction model
 }
 \author{
 Hiromi Otsuka
 }
 \address{
 Department of Physics, Tokyo Metropolitan University, Tokyo 192-0397,
 Japan
 }
\date{\today}
\begin{abstract}

 The Berezinskii-Kosterlitz-Thouless-type continuous phase transition
 observed in the three-spin interaction model is discussed.
 The relevant field theory describes the topological defects involved
 and enables us to perform the renormalization-group analysis.
 Based on it,
 we shall propose the finite-size-scaling ansatz for the helicity
 modulus which exhibits the exponent $\bar\nu=3/5$ for the correlation
 length in the disordered phase.
 We perform the Monte Carlo simulations to confirm the ansatz.
 Also, we argue its relevance to the ground-state phase transition
 in the quantum spin chain.

 \end{abstract}
 \maketitle

 \section{INTRODUCTION}

 The universality observed in the phase
 transitions is one of the most important phenomena to understand the
 interaction effects.
 For the two-dimensional (2D) critical systems other than those with
 intrinsic anisotropy,
 it is pronouncedly expressed by the conformal symmetry, and the
 corresponding field theories are characterized by the central charge
 $c$
 \cite{BPZ}.
 In the case $c<1$,
 it appears to almost specify the universality class, i.e., the possible
 set of the critical exponents
 \cite{FQS}.
 However, for larger values of $c$,
 there still exist considerable efforts to understand their
 universalities, which are possibly related to the exotic phase
 transitions observed in the complicated systems.
 It is widely known that the frustration effects sometimes bring about
 the critical ground states as well as the
 finite-temperature critical points with larger values of $c\ge1$.
 Thus, they have been gathering great attention over the years.
 On another front,
 the multispin interactions appear to include the same effect:
 The exactly solved Baxter-Wu model consisting of the three-Ising-spin
 product interaction is the most basic one
 \cite{Baxt73},
 which belongs to the same universality class as the four-state Potts
 ferromagnet
 \cite{Alca99}.
 The Ising and the four-state Potts criticalities are of $c=1/2$ and 1,
 respectively.
 Thus, the multispin interactions can be expected as another source
 to bring about the larger value of $c$.
 
 In this paper, we investigate the three-spin interaction model (TSIM)
 introduced by Alcaraz et al
 \cite{Alca83}.
 Suppose that $\langle k,l,m\rangle$ denotes three sites at the corners
 of each elementary plaquette of the triangular lattice
 $\Lambda$
 (which consists of three sublattices
 $\Lambda_{\rm a}$, $\Lambda_{\rm b}$, and $\Lambda_{\rm c}$),
 then the following reduced Hamiltonian expresses a class of TSIM:
 \begin{eqnarray}
  {\beta H}
   =-\frac{J}{k_{\rm B}T}\sum_{\langle k,l,m\rangle}
   \cos\left(\varphi_k+\varphi_l+\varphi_m\right). 
   \label{eq_Hamil}
 \end{eqnarray}
 The angle variables
 $\varphi_k\in[0,2\pi)$
 are located on sites, and the model parameter, the temperature $T$,
 will be measured in units of $J/k_{\rm B}$.
 In the previous paper
 \cite{Otsu07},
 we discussed an effective field theory for a related model (i.e., its
 clock version) and predicted the one with $c=2$, which was followed by
 the numerical confirmation based on the transfer-matrix calculations.
 Here, we perform the renormalization-group (RG) analysis to
 predict the Berezinskii-Kosterlitz-Thouless (BKT)
 \cite{Bere71,Kost73Kost74}
 type phase transition between the critical and the disordered phases. 
 In particular,
 we shall propose the finite-size-scaling (FSS) ansatz for the helicity
 modulus representing the stiffness of the corresponding interface model
 \cite{Fish73,Kost77Ohta79}.
 The discussion goes on in a parallel way with the $c=1$ BKT transition
 case
 \cite{HaradaKawashima},
 and the ansatz includes the exponent $\bar\nu$ for the correlation
 length. 
 However,
 since the topological defects involved
 (for its definition, see below)
 are not described by the scalars (vortexes), but by the vectors
 \cite{Alca83}
 like the Burgers vectors for the dislocations in the 2D melting
 \cite{Halp78Nels78Nels79,Youn79},
 some modifications occur in the RG analysis, and then bring about
 $\bar\nu=3/5$.
 The FSS ansatz permits us to check it directly
 by the numerical method,
 and then provides solid evidence to support the theoretical
 prediction on the instability in the $c=2$ criticality.

 \section{THEORY}

 The fact that the system is invariant under the global spin rotations,
 $\varphi_k\to
   \varphi_k+
   \sum_{\rho={\rm a,b,c}}
   \sum_{l\in\Lambda_\rho}
   \psi_\rho\delta_{k,l}$,
 with the condition
 $\psi_{\rm a}+\psi_{\rm b}+\psi_{\rm c}=0$ (mod $2\pi$), 
 is important. 
 Although this U(1)$\times$U(1) symmetry defined by two independent
 phases, say $\psi_{\rm a,b}$, is not broken and gives the
 low-temperature critical phase, it specifies the possible type of
 perturbations to bring about the high-temperature disordered phase.
 According to Ref.\ [\onlinecite{Otsu07}],
 the following vector sine-Gordon Lagrangian density is relevant to the
 effective description:
 \begin{align}
  {\cal L}
  =
  \frac{K}{4\pi}
  \|\partial_i{\bf\Phi}({\bf x})\|^2
  +
  \frac{y_{1}}{2\pi a^2}
  \sum_{~\|{\bf N}\|=1~~}:{\rm e^{i{\bf N\cdot\Theta({\bf x})}}}:.
  \label{eq_L}
 \end{align}
 The symbol :~: denotes the normal ordering and means the subtraction of
 contractions of fields between them; $\|~\|$ indicates the norm of
 the vector, and $a$ gives a short-distance cutoff. 
 The parameter $y_1$ stands for the effective fugacity to control the
 appearance of topological defects
 \cite{Otsu07}.
 ${\bf\Phi}({\bf x})$ is the two component vector field attached at the
 position ${\bf x}$ in the basal
 2D space,
 so the first term represents
 the interface model,
 where $K$ gives its stiffness.
 The above symmetry is realized as the periodicity of the field, i.e.,
 ${\bf\Phi}\equiv{\bf\Phi}+2\pi{\bf e}_\alpha$ $(\alpha=1,2)$, where
 ${\bf e}_{1,2}$ are the normalized fundamental vectors of the so-called
 repeat lattice ${\cal R}$
 \cite{Kond95Kond96}
 isomorphic to the triangular lattice
 \cite{Otsu07}. 
 Then,
 the interface can acquire the discontinuity of the amount
 $2\pi{\bf N}$ with ${\bf N}\in{\cal R}$.
 The second term consists of
 the vertex
 operators where
 ${\rm i}K\partial_i{\bf \Phi}=\epsilon_{ij}\partial_j{\bf \Theta}$,
 and creates the shortest discontinuities
 among possible ones  (i.e., the length $\|{\bf N}\|=1$).
 From the RG viewpoint, these topological defects controlled by $y_1$
 are enough to be kept in the theory because these are the most relevant
 ones (see below).

 The reason why we have started with the Lagrangian density instead of
 the vector Coulomb gas (CG) representation
 \cite{Alca83}
 is that, for the RG analysis we shall employ the conformal field theory
 (CFT) technology, which requires the scaling dimensions of local density
 operators and the operator-product-expansion (OPE) coefficients among
 them.
 While details of their derivations on the Gaussian fixed point (i.e.,
 the first term) will be given in our future report
 \cite{Otsu08},
 here we shall summarize the relevant results to the present case.
 Since the low-temperature critical phase corresponds to the Gaussian
 fixed line parameterized by $K$, the so-called ${\cal M}$ operator
 \cite{Kada79Kada79B},
 \begin{align}
  {\cal M}({\bf x})
  \equiv
  \frac{Ka^2}{\sqrt8}\|\partial_i{\bf\Phi}({\bf x})\|^2,
  \label{eq_M_op}
 \end{align}
 which shifts the system along the line, is the most important one.
 Since
 $\langle{\cal M}({\bf x}){\cal M}({\bf 0})\rangle_0=\left(a/r\right)^4$
 independently of $K$
 ($r$ is the distance between ${\bf 0}$ and {\bf x}), it is truly
 marginal.
 On the other hand,
 the normalized form of the second term in Eq.\ (\ref{eq_L}) is given as
 \begin{align}
  {\cal W}({\bf x})
  &\equiv
  \frac{1}{\sqrt6}\sum_{\|{\bf N}\|= 1}
  :{\rm e^{i{\bf N\cdot\Theta}({\bf x})}}:, 
  \label{eq_W}
 \end{align}
 whose dimension is $x_{\cal W}\equiv K\|{\bf N}\|^2/2=K/2$.
 Thus, ${\cal W}$ becomes marginal at $K=K^*$ $(\equiv4)$ and brings
 about the transition to the disordered phase for $K<K^*$.
 For later RG argument, the expansion of the operator product
 ${\cal W}({\bf x}){\cal W}({\bf 0})$ is crucial.
 While there are 36 terms in the double summations with respect to the
 vector charges
 (say ${\bf N}$ and ${\bf N'}$),
 the following two cases are enough to be taken into account:
 (i) ${\bf N+N'=0}$
 and 
 (ii) $\|{\bf N+N'}\|=1$ (the other terms are irrelevant here).
 After some calculus using the complex coordinate and employing the
 chiral decomposed form of fields as usual,
 we find that the cases (i) and (ii) mainly give
 ${\cal M}$ and ${\cal W}$, respectively.
 Then, the expression of the OPE becomes as follows
 \cite{Otsu08,Polchinski}:
 \begin{align}
 &{\cal W}({\bf x}){\cal W}({\bf 0})\nonumber\\
 &\simeq
  \frac{x_{\cal W}}{\sqrt2}
  \left(\frac{a}{r}\right)^{2x_{\cal W}-2}\!\!{\cal M}({\bf 0})
  +
  \frac{2}{\sqrt6}
  \left(\frac{a}{r}\right)^{ x_{\cal W}  }\!\!{\cal W}({\bf 0})
  +
  \cdots,
 \end{align}
 where ``$\cdots$'' includes the unit operator and the stress tensor as
 well as less singular terms.
 Consequently, we can obtain the OPE coefficients
 $C_{\cal WWM}=x_{\cal W}/\sqrt2$ and
 $C_{\cal WWW}=2/\sqrt6$.
 The significance of this result is as follows:
 Due to the triangular lattice structure of ${\cal R}$,
 three vector charges at the angle of 120 degrees to each other
 (e.g., ${\bf e}_1$, $-{\bf e}_1+{\bf e}_2$, and $-{\bf e}_2$)
 satisfy the vector charge neutrality condition, and then provide the
 nonzero value of $C_{\cal WWW}$.
 This brings about the difference from the BKT transition
 (see below).

 Now, the RG equations are derived as follows. 
 Suppose that
 the critical fixed point is perturbed by the
 marginal scalar operators
 ${\cal O}_\mu({\bf x})$ (normalized) as 
 ${\cal L}={\cal L}_{0}^*+\sum_\mu \lambda_\mu{\cal O}_\mu/2\pi a^2$,
 then, the RG equations for the change of the cutoff
 $a\to(1+{\rm d}l)a$
 are given by
 ${\rm d}\lambda_\mu/{\rm d}l
 =-(1/2)\sum_{\nu,\rho}C^*_{\mu\nu\rho}\lambda_\nu\lambda_\rho$
 within the one-loop calculations
 \cite{Poly72}.
 For the present case of the couplings $y_0$ ($\equiv K/K^*-1$) and $y_1$,
 the equations become
 \begin{align}
  \frac{{\rm d}y_0(l)}{{\rm d}l}
  =
  -3y_1(l)^2,
  ~
  \frac{{\rm d}y_1(l)}{{\rm d}l}
  =
  -2y_0(l)y_1(l)-y_1(l)^2.
  \label{eq_RG_TH}
 \end{align}
 Whereas these are similar to the BKT RG equations
 \cite{Bere71,Kost73Kost74},
 the $y_1^2$ term emerges in the second equation due to the nonvanishing
 coefficient $C_{\cal WWW}$ (see also
 \cite{Alca83}).
 \ZUichi
 Now, we shall discuss its consequences. 
 Like to the BKT transition case,
 these equations possess a conserved quantity,
 but unlike to the case,
 it is given as a homogeneous expression of degree five in $y_0$ and
 $y_1$ as
 \begin{align}
  \Delta^3
  \equiv
  (y_1-y_0)^3 \Bigl(y_1+\frac23y_0\Bigr)^2~~~(\Delta\in{\mathbb R}).
  \label{ConservedQuantity}
 \end{align}
 Figure\ \ref{contour-y0-y1}(a)
 gives the contour plot with arrows to show the trajectories of the RG
 flows.
 Then,
 we see that the line $y_1=y_0$ is the separatrix between the critical
 and the disordered phases, so the small parameter for the phase
 transition can be introduced as $y_1(0)=(1+t)y_0(0)$, with $|t|\ll1$.
 Now,
 with the aid of the conservation law, we can analyze the RG flows and
 obtain the following form:
 \begin{align}
  y_{0,1}(l)=|\Delta|^{\frac35}g_{0,1}^\pm(|\Delta|^{\frac35}l+\tau_t),
  \label{eq_form_pm}
 \end{align}
 where the superscript ``$\pm$'' indicates the sign of $\Delta$ and
 $\tau_t$ is a certain function of $t$ given by the initial values of
 the couplings
 \cite{Youn81,Minn87}.
 This form exhibits a kind of self-similarity
 that the trajectories with the initial conditions having the same
 values of $t$
 [e.g., all points on the red dotted line in Fig.\ \ref{contour-y0-y1}(a)]
 fall into a single curve independently of $\Delta$
 [the red dotted line (a numerical integration) in Fig.\
 \ref{contour-y0-y1}(b) or \ref{contour-y0-y1}(c)],
 and
 that the $t$ dependence can be absorbed by changing the origin of the
 scaled variable according to $\tau_t$ (the red dotted line overlaps
 with the pink short-dashed line).
 The explicit forms of $g_{0,1}^\pm$ are complicated while those
 in the BKT transition are the trigonometric or the hyperbolic
 functions
 \cite{Youn81,Minn87}.
 Nevertheless,
 we can extract some properties because they share the basic feature
 with the BKT transition case.
 On the separatrix $t=0$, the solution is simply given by
 $y_{0,1}(l)=y(l)\equiv1/3(l+l_{T_{\rm c}})$
 with
 $l_{T_{\rm c}}\equiv 1/3y(0)$.
 Around it,
 we find
  $y_{0,1}(l)\simeq y(l)-[(1\pm2)/4](3/5)^{2/3}\Delta/y^{2/3}(l)$
 when
 $|t|^{3/5}(l/l_{T_{\rm c}})\ll1$
 is satisfied
 (the upper sign refers to the former).
 Note
 that these expressions exhibit the nonsingularities of $y_{0,1}(l)$ at
 $\Delta=0$ and also that they can be regarded as the expansions of the
 following scaling forms:
 \begin{align}
  y_{0,1}(l)\simeq
 (l+l_{T_{\rm c}})^{-1}\Psi_{0,1}[\Delta(l+l_{T_{\rm c}})^{\frac53}]
  \label{eq_form_1}
 \end{align}
 with $\Psi_{0,1}(X)\simeq1/3+O(X)$.
 To cast these into the forms compatible with Eq.\ (\ref{eq_form_pm}), 
 we should further replace $l_{T_{\rm c}}$ by a temperature dependent
 parameter $l_T\equiv \tau_t/|\Delta|^{3/5}$. 
 However,
 since $l_T$ means the logarithmic scale to obtain the initial values
 $y_{0,1}(0)$ from $y_{0,1}\simeq\infty$ lying almost on the separatrix,
 it is a smooth function and can be represented by its value at the
 transition point, $l_{T_{\rm c}}$, near $T_{\rm c}$.

 Now, we shall estimate the correlation length
 from Eqs.\ (\ref{eq_form_pm}) and (\ref{eq_form_1}).
 When we write the characteristic logarithmic scale as $l_{\rm F}$
 satisfying the condition
 $y_0(l_{\rm F})=-\infty$,
 then
 $\xi(T)\propto {\rm exp}(l_{\rm F})$.
 Figure\ \ref{contour-y0-y1}(b) exhibits
 $g_0^+(\tau\simeq1.4)=-\infty$,
 and
 $\Delta$ is proportional to
 $T-T_{\rm c}$
 near
 $T_{\rm c}$.
 Thus,
 we obtain
 $\xi(T)\propto {\rm exp}[\text{const}/(T-T_{\rm c})^{\bar\nu}]$
 with the exponent
 $\bar\nu=3/5$.
 Unlike to the BKT transition case $\bar\nu=1/2$
 \cite{Bere71,Kost73Kost74},
 our theory predicts the above exponent, which is attributed to the
 nature of the topological defects characterized by the vector charges,
 and more precisely to the nonvanishing OPE coefficient $C_{\cal WWW}$.
 However,
 our result also disagrees with the previous one based on the vector CG
 representation
 \cite{Alca83}.
 Therefore,
 the numerical evidence to support our theory is desired.

 In the remainder of the paper,
 we shall provide the evidence by the use of the Monte Carlo (MC) method.
 For this purpose,
 here we explain the FSS property of the helicity modulus to focus on
 below.
 The helicity modulus is defined as the response of the free energy
 against the long-wave-length ($\lambda$) twist of the local order
 field, which is proportional to the square of the wave number
 $q\equiv2\pi/\lambda$, i.e., 
 $\Upsilon(T)\equiv\lim_{q\to0}\partial^2 f(T,q)/\partial q^2$
 \cite{Fish73,Kost77Ohta79}. 
 For the present model with the U(1)$\times$U(1) symmetry,
 the twist can be imposed by, for instance,
 $\forall~l\in \Lambda_{\rm b};~\varphi_l \to \varphi_l+{\bf q}\cdot{\bf x}_l$
 and
 $\forall~m\in \Lambda_{\rm c};~\varphi_m \to \varphi_m-{\bf q}\cdot{\bf x}_m$,
 where
 ${\bf x}_l$ (${\bf x}_m$) is the position vector of the $l$th ($m$th) site
 and
 ${\bf \hat{q}}\equiv{\bf q}/q$ is a unit vector in the $x$ direction of
 the basal 2D space.
 Then,
 one finds the following expression:
 \begin{align}
  \Omega\Upsilon(T)
  =
  \Bigl\langle
  -\frac12 H
  -\frac1T
  \Bigl(
  \sum_{\langle klm\rangle}\sin\varphi_t~{\bf \hat{q}}\cdot{\bf x}_{lm}
  \Bigr)^2
  \Bigr\rangle, 
 \end{align}
 where
 $\varphi_t\equiv\varphi_k+\varphi_l+\varphi_m$, 
 ${\bf x}_{lm}\equiv{\bf x}_l-{\bf x}_m$, and
 $\Omega$ is the 2D volume of the system.
 While this is useful for numerical estimations of $\Upsilon$, 
 its theoretical expression is also obtained as follows.
 According to Ref.\
 [\onlinecite{Otsu07}],
 the above twist corresponds to the following shift in the vector field: 
 $
 {\bf\Phi}({\bf x})\to
 {\bf\Phi}({\bf x})+{\bf q}\cdot{\bf x}(-{\bf e}_1+{\bf e}_2)
 $,
 which then brings about the increase of the Gaussian part in the free
 energy density $(TK/2\pi)(q^2/2)$. 
 For $T\le T_{\rm c}$,
 we can renormalize the effects of the topological defects,
 so that the helicity modulus in the full theory is given by
 $\Upsilon(T)=TK(l=\infty)/2\pi$.
 As usual,
 the ratio $\pi\Upsilon(T)/2T$ exhibits the universal jump from 1 to 0
 at $T=T_{\rm c}$ in the thermodynamic limit
 although it is rounded off in the finite-size systems.
 Relating to this,
 Harada and Kawashima performed the MC simulations of the 2D quantum XY
 model, and gave the FSS analysis of the helicity modulus to confirm the
 BKT theory being valid there
 \cite{HaradaKawashima}.
 According to their discussion,
 here we propose the FSS form for the present phase transition:
 For the finite-size systems with
 the linear dimension $L$ near $T_{\rm c}$, 
 $l+l_{T_{\rm c}}$ and $\Delta$ in Eq.\ (\ref{eq_form_1}) are replaced
 by the logarithmic scale $\ln(L/L_0)$  ($L_0$ could depend on the
 temperature) and $T-T_{\rm c}$, respectively. 
 Then, we obtain
 \begin{align}
 \frac{\pi\Upsilon(T,L)}{2T}-1
 =
 \frac{\Psi_0\{(T-T_{\rm c})[\ln({L}/{L_0})]^{\frac53}\}}{\ln({L}/{L_0})},
 \label{eq_FSS}
 \end{align}
 where $T_{\rm c}$ and $L_0$ are the parameters to be determined based
 on the FSS ansatz.

 \section{NUMERICAL CALCULATIONS}

 \ZUni

 \ZUsan

 Here, we shall provide the MC data of
 the helicity modulus and its FSS analysis (the details of our
 simulations will be given elsewhere).
 Using the standard Metropolis algorithm, the systems of sizes up to
 $L=384$ were treated.
 The $10^7$ MC steps (MCS) were used for samplings after the $10^6$ MCS
 for equilibration. 
 At each temperature,
 we performed the independent runs up to 64 so as to attain reliable
 statistics of data.
 In units of
 the squared lattice constant $a^2=1$,
 the volume is given as $\Omega=L^2/\zeta$, where $\zeta$ ($=2/\sqrt3$) is
 the geometric factor of $\Lambda$.
 Then,
 we obtain the helicity modulus given in Fig.\ \ref{HelicityModulus}.
 As expected, it exhibits the steep decrease
 and crosses the straight line $2T/\pi$ (see the circles for $L=384$).
 Since $\Upsilon$ stands for the stiffness of the interface, its
 reduction to 0 indicates a kind of melting.
 The inset gives the magnified view near the transition temperature,
 where ${\pi\Upsilon}/{2T}$ is plotted against $T$.
 Thus, the crossings with the dotted line provide the finite-size
 estimates of $T_{\rm c}$, $T_{\rm c}(L)$.
 In Fig.\ \ref{scaling},
 we show the FSS plot of $\Upsilon$.
 According to Eq.\ (\ref{eq_FSS}),
 the parameters $T_{\rm c}$ and $L_0$ were searched until the best
 collapsing of the scaled data is achieved.
 Actually,
 we performed it by eye and obtained the values, 
 $T_{\rm c}\simeq\TcScaling$ and $L_0\simeq\L0Scaling$. 
 Then,
 the figure exhibits that the scaling region is narrow in the upper side
 of $T_{\rm c}$: 
 For $L\ge192$, it continues up to about 0.25 in the
 abscissa (see the diamonds and circles).
 Although this is due to the limitation on the size treated, and
 further to the property of $\Upsilon$ showing the jump in the
 thermodynamic limit, we can check the reliability of our FSS analysis
 as follows.
 As in the case of the BKT transition,
 we can extrapolate the finite-size estimates as
 $T_{\rm c}(L)=T_{\rm c}(\infty)+c_1/(\ln L+c_2)^{5/3}$,
 where $T_{\rm c}(\infty)$ and $c_{1,2}$ are the least-squares-fitting
 parameters.
 Then,
 we obtain
 $T_{\rm c}(\infty)\simeq\TcExtrap$, which agrees well with the FSS
 result.
 Further,
 as argued, the scaling function is theoretically expected
 to satisfy $\Psi_0(0)=1/3$. 
 We denote this condition by the cross in Fig.\ \ref{scaling}
 and then we find the good coincidence with the FSS analysis.
 With respect to the possibility of $\bar\nu\ne3/5$, we tried the FSS
 analysis with its value, for instance, 2/5, but the goodness of the
 scaling in Fig.\ \ref{scaling} could not be reproduced within our
 search.
 Consequently,
 these numerical data and the consistencies among them strongly support
 our field theoretical description on the BKT-type phase transition
 observed in TSIM \cite{COMM0}.

 \section{DISCUSSIONS AND SUMMARY}

 We shall mention the relevance of our
 theory to other systems.
 The bilinear-biquadratic spin-1 chain is exactly solvable in some cases.
 The Uimin-Lai-Sutherland (ULS) point is one of them, on which the
 low-energy excitations are described by the level-1 SU(3)
 Wess-Zumino-Witten model ($c=2$), and by which the massless quadrupole
 and the massive Haldane phases are separated
 \cite{Uimi70Lai74Suth75}.
 This ground-state phase transition was clarified to be BKT-type, and in
 terms of the small parameter to control the distance from the ULS point,
 the correlation length
 is given in the identical
 form as the present one (i.e., $\bar{\nu}=3/5$)
 \cite{Itoi97}.
 Therefore,
 we think that the phase transition discussed here may share the same
 fixed-point properties with it;
 we shall discuss this issue more closely in our future study
 \cite{Otsu08}.
 To summarize,
 based on the vector sine-Gordon field theory, we have discussed the
 continuous phase transition observed in TSIM; the FSS ansatz for the
 helicity modulus was proposed based on the RG analysis.
 Our theoretical predictions were confirmed by the use of the MC
 simulations.

 The author thanks
 S. Hayakawa,
 A. Tanaka,
 Y. Okabe,
 N. Kawashima, 
 and 
 K. Nomura
 for stimulating discussions. 
 This work was supported by
 Grants-in-Aid from the Japan Society for the Promotion of Science,
 Scientific Research (C), Grant No.\ 17540360.

 \newcommand{\AxS}[1]{#1,}
 \newcommand{\AxD}[2]{#1 and #2,}
 \newcommand{\AxT}[3]{#1, #2, and #3,}
 \newcommand{\AxQ}[4]{#1, #2, #3, and #4,}
 \newcommand{\REF }[4]{#1 {\bf #2}, #3 (#4)}
 \newcommand{\JPSJ}[3]{\REF{J. Phys. Soc. Jpn.\           }{#1}{#2}{#3}}
 \newcommand{\PRL }[3]{\REF{Phys. Rev. Lett.\             }{#1}{#2}{#3}}
 \newcommand{\PRA }[3]{\REF{Phys. Rev.\                  A}{#1}{#2}{#3}}
 \newcommand{\PRB }[3]{\REF{Phys. Rev.\                  B}{#1}{#2}{#3}}
 \newcommand{\PRE }[3]{\REF{Phys. Rev.\                  E}{#1}{#2}{#3}}
 \newcommand{\NPB }[3]{\REF{Nucl. Phys.\                 B}{#1}{#2}{#3}}
 \newcommand{\JPA }[3]{\REF{J. Phys.\ A                   }{#1}{#2}{#3}}
 \newcommand{\JPC }[3]{\REF{J. Phys.\ C: Solid State Phys.}{#1}{#2}{#3}}


\begin{thebibliography}{99}
\bibitem{COMM0}
 In Ref.\ \cite{Otsu07},
 the explanation that $\bar\nu$ takes 2/5 was given. 
 But, this should be corrected as $\bar\nu=3/5$.
\end{thebibliography}
\end{document}